\documentclass[11pt]{article}

\usepackage{html,amsmath,amssymb,amsfonts,epsfig,cancel,cite,longtable,caption,siunitx,array,color,booktabs,setspace}

\usepackage{dsfont}

\renewcommand\baselinestretch{1.23}
\numberwithin{equation}{section}

\newcommand{\be}{\begin{equation}}
\newcommand{\ee}{\end{equation}}
\newcommand{\beq}{\begin{equation}}
\newcommand{\eeq}{\end{equation}}
\newcommand{\ba}{\begin{array}}
\newcommand{\ea}{\end{array}}
\newcommand{\bea}{\begin{eqnarray}}
\newcommand{\eea}{\end{eqnarray}}
\newcommand{\bean}{\begin{eqnarray*}}
\newcommand{\eean}{\end{eqnarray*}}

\newcommand{\nn}{\nonumber}

\newcommand{\rar}{\rightarrow}

\newcommand{\rk}{\mathop{{\rm rk}}}

\newcommand{\ch}{{\rm ch}}

\newcommand{\ind}{\,\mathop{{\rm Ind}}}

\newcommand{\cO}{{\cal O}}
\newcommand{\cN}{{\cal N}}

\newcommand{\cA}{{\cal A}}
\newcommand{\cB}{{\cal B}}
\newcommand{\cC}{{\cal C}}

\newcommand{\cV}{{\cal V}}

\newcommand{\fn}{\footnotesize}

\def\cjn1{{\cA, \cC^*\otimes \wedge^j \cN^*}}
\def\bjn1{{\cA, \cB^*\otimes \wedge^j \cN^*}}
\def\vjn1{{\cA, \cV^*\otimes \wedge^j \cN^*}}
\def\cjn2{{\cA, \cC\otimes \wedge^j \cN^*}}
\def\bjn2{{\cA, \cB\otimes \wedge^j \cN^*}}
\def\vjn2{{\cA, \cV\otimes \wedge^j \cN^*}}

\newcommand{\varstr}[2]{\vrule height #1 depth #2 width0pt}

\begin{document}

\title{{\LARGE \bf$~$\\[-7pt]
A Heterotic Standard Model\\[9pt] 
with $B-L$ Symmetry and a Stable Proton \\ [9pt]
}}

\vspace{3cm}

\author{
Evgeny I. Buchbinder${}^{1}$,
Andrei Constantin${}^{2}$,
Andre Lukas${}^{2}$
}
\date{}
\maketitle
\thispagestyle{empty}
\begin{center} { ${}^1${\it The University of Western Australia, \\
35 Stirling Highway, Crawley WA 6009, Australia\\[0.3cm]
       ${}^2$Rudolf Peierls Centre for Theoretical Physics, Oxford
       University,\\
       $~~~~~$ 1 Keble Road, Oxford, OX1 3NP, U.K.}}\\
       
\end{center}

\vspace{11pt}
\abstract
\noindent We consider heterotic  Calabi-Yau compactifications with $S(U(4) \times U(1))$ background gauge fields. These models lead to gauge groups with an additional $U(1)$ factor which, under certain conditions, can combine with hypercharge to a $B-L$ symmetry. The associated gauge boson is automatically super-massive and, hence, does not constitute a phenomenological problem. We illustrate this class of compactifications with a model based on the monad construction, which leads to a supersymmetric standard model with three families of quarks and leptons, one pair of Higgs doublets, three right-handed neutrinos and no exotics charged under the standard model group. The presence of the $B-L$ symmetry means that the model is safe from proton decay induced by dimension four operators. Due to the presence of a special locus in moduli space where the bundle structure group is Abelian and the low-energy symmetry enhances we can also show the absence of dimension five proton-decay inducing operators.

\vskip 5cm
{\hbox to 7cm{\hrulefill}}
\noindent{\fn evgeny.buchbinder@uwa.edu.au}\\
{\fn a.constantin1@physics.ox.ac.uk}\\
{\fn lukas@physics.ox.ac.uk} 
\newpage

\tableofcontents

%
%
\section{Introduction}


Finding an ultraviolet completion of the standard model of particle physics within string theory has been one of the main quests in theoretical high energy physics in the last few decades. It has been understood some time ago that Calabi-Yau compactifications of the $E_8\times E_8$ heterotic string~\cite{Candelas:1985en}  provide a promising avenue towards realistic particle phenomenology~\cite{Greene:1986bm,Greene:1986jb}. However, finding concrete models with the desired phenomenological properties has been a difficult task. Until recently, only few examples of heterotic Calabi-Yau compactifications exhibiting the particle content of the supersymmetric standard model with no exotics existed in the literature~\cite{Bouchard:2005ag, Braun:2005ux, Braun:2005bw, Anderson:2009mh,Braun:2011ni}. More recently, large classes of such models, based on constructions with Abelian vector bundles, have been found~\cite{Anderson:2011ns, Anderson:2012yf,Anderson:2013xka}, and these open up the possibility to study more detailed phenomenology beyond the particle spectrum.

At the next level, one needs to obtain suitable couplings, in order to avoid well-known problems encountered in supersymmetric extensions of the Standard Model and conventional four dimensional GUTs, such as fast proton decay. In the MSSM, an additional symmetry is required to forbid operators which can lead to fast proton decay and frequently a $B-L$ symmetry or a discrete sub-group thereof is being used for this purpose.
In the context of the heterotic string, standard models with a $B-L$ symmetry have been realized in Refs.~\cite{Braun:2005ux, Braun:2005bw, Anderson:2009mh}. However, in these models $B-L$ is a local symmetry and the associated vector boson is massless at the string level. It has been shown that, under certain conditions, such a local $B-L$ symmetries can be broken spontaneously well below the string scale but above the electroweak scale~\cite{Ambroso:2009jd, Ambroso:2009sc}, as a result of renormalization group running.

In the present paper we take a different approach to constructing a heterotic model with $B-L$ symmetry. Compactifications of the $E_8\times E_8$ heterotic string with $S\left(U(4)\times U(1) \right)$-bundles lead to GUT models with gauge group $SU(5)\times U(1)$. Related compactifications have previously been studied in Refs.~\cite{Distler:1987ee,Blumenhagen:2005ga,Blumenhagen:2006ux,Blumenhagen:2006wj}. The additional $U(1)$ is generically Green-Schwarz anomalous and, consequently, the corresponding vector boson receives a super-heavy St\"uckelberg mass. Hence, below the string scale, the additional $U(1)$ symmetry is effectively global. Under certain conditions, this global $U(1)$ symmetry, combined with hypercharge, leads to a $B-L$ symmetry, which prohibits the presence of dangerous dimension 4 operators. 

In a previous publication \cite{Buchbinder:2013dna}, we have been led to considering such compactifications by the study of non-Abelian deformations of holomorphic line bundle sums, based on a particular example from the data base of heterotic line bundle standard models in Ref.~\cite{Anderson:2011ns, Anderson:2012yf}. The inverse process, the splitting of bundles at stability walls, has been in general described in Refs.~\cite{Anderson:2009sw,Anderson:2009nt}.  For the particular model, we have shown in Ref.~\cite{Buchbinder:2013dna} that there exists a locus in moduli space where the bundle structure group generically is $S\left(U(4)\times U(1)\right)$ and the low-energy gauge group is $SU(5)\times U(1)$. The additional $U(1)$ indeed leads to a $B-L$ symmetry so that the model is safe from proton decay induced by dimension four operators. More significantly, the presence of a special locus with Abelian structure group, which implies symmetry enhancement to $SU(5)\times S(U(1)^5)$ in the low-energy theory, leads to additional control over the coupling constants in the theory. The relevance of enhanced symmetry in bundle moduli space in the context of model building has been pointed out in Ref.~\cite{Kuriyama:2008pv,Anderson:2010tc}. For our model, the presence of this locus implies the absence of dimension five operators inducing proton decay.  In Ref.~\cite{Buchbinder:2013dna}, this model has only been worked out in detail at the level of a GUT with gauge group $SU(5)\times U(1)$.

The main purpose of the present paper, is to construct the associated standard model and show that it has indeed all the features anticipated from the associated GUT. In this way we are led to a model which enjoys a number of phenomenologically attractive properties:
\begin{itemize}
\item A standard model gauge group and an MSSM spectrum with three generations of quarks and leptons, one pair of Higgs doublets, three right-handed neutrinos plus a number of moduli uncharged under the standard model group.
\item The absence of any exotic particles charged under the standard model group; in particular, Higgs triplets are projected out by the Wilson lines. 
\item The presence of a global $B-L$ symmetry.
\item The absence of dimension 4 and dimension 5 operators which could trigger a fast proton decay.
\end{itemize}
The paper is organised as follows. In Section~\ref{sec:general} we discuss some general aspects of heterotic compactifications with $S\left(U(4) \times U(1)\right)$-bundles. In particular, we distinguish regular models where the additional $U(1)$ leads to a $B-L$ symmetry and irregular models where the additional $U(1)$ symmetry is different from $B-L$. In Section~\ref{sec:regmodels} we focus on regular models. The structure of our specific model, at the GUT level, is reviewed in Section~\ref{sec:GUTmodel} and the associated standard model is constructed in Section~\ref{sec:SM}. Proton stability for this model is discussed in Section~\ref{sec:proton}. Some technical details are collected in two Appendices. Appendix A presents a number of useful formulae relevant to bundles with $S(U(4)\times U(1))$ structure group and Appendix B outlines the calculation of the Higgs spectrum of the model.

Most of the technical computations presented in this paper were done using the ``CICY package" (described in~\cite{Anderson:2007nc, Gray:2007yq, Anderson:2008uw, He:2009wi, Anderson:2009mh}) and the database of line bundle models~\cite{Linebundles}. 


\section{Heterotic compactifications with $S\left(U(4)\times U(1)\right)$-bundles}
\label{sec:general}
Let us start by discussing in full generality heterotic $E_8\times E_8$ compactifications on a smooth Calabi-Yau three-fold $X$ carrying a holomorphic vector bundle $V$ with $S(U(4)\times U(1))$ structure group. Such bundles can be written as a Whitney sum
\begin{equation}
 V=U\oplus L\; , \label{Vdef}
\end{equation} 
where $L$ is a line bundle and $U$ is a rank four bundle with $U(4)$ structure group satisfying ${\rm c}_1(U)=-{\rm c}_1(L)$, so that ${\rm c}_1(V)=0$. As usual, we demand that
\begin{equation}
{\rm c}_2(TX) - {\rm c}_2(V) = {\rm c}_2(TX) - {\rm c}_2(U)+\frac{1}{2}{\rm c}_1(L)^2\stackrel{!}{\in} \text{Mori cone of }X
\end{equation}
in order to be able to satisfy the heterotic anomaly cancellation condition. For the bundle $V$ to preserve supersymmetry it needs to be poly-stable with slope zero.  In particular, for a bundle with splitting type as in Eq.~\eqref{Vdef} the slope of $L$ must vanish, that is,
\begin{equation}
 \mu(L)=\int_X {\rm c}_1(L)\wedge J\wedge J\stackrel{!}{=}0\; ,
\end{equation}
where $J$ is the K\"ahler form on $X$. This equation amounts to a constraint on the K\"ahler parameters and effectively constraints the models to a co-dimension one locus in K\"ahler moduli space. Note, since ${\rm c}_1(U)=-{\rm c}_1(L)$, it follows that the slope of $U$ also vanishes. In addition, poly-stability of $V$ requires $U$ to be stable, that is all sub-sheafs ${\cal F}\subset U$ with $0<{\rm rk}({\cal F})<4$ must satisfy $\mu({\cal F})<0$. 

To discuss the structure of the low-energy theory, we begin by looking at the group theory relevant for models of this type. The embedding of the structure group into the observable $E_8$ can be realised via the two sub-group chains:
\begin{eqnarray}
 E_8&\rar& SU(4)\times SO(10)\rar SU(4)\times U_2(1)\times SU(5)\nn\\
       &\rar& SU(4)\times U_2(1)\times SU_c(3)\times SU_W(2)\times U_1(1)\label{decomp1}\\
 E_8&\rar&SU(5)\times SU(5)\rar SU(4)\times U_2'(1)\times SU(5)\nn\\
       &\rar&SU(4)\times U_2'(1)\times SU_c(3)\times SU_W(2)\times U_1(1)\; .  \label{decomp2}    
\end{eqnarray} 
Note that the $U_2(1)$ in the first decomposition arises from $SO(10)\rar U_2(1)\times SU(5)$ while the $U_2'(1)$ symmetry in the second decomposition is due to $SU(5)\rar SU(4)\times U_2'(1)$. All charge normalizations and the subsequent branchings are taken from Ref.~\cite{Slansky:1981yr}.
Under the first chain of sub-groups, the fundamental representation ${\bf 248}_{E_8}$ of $E_8$ branches as
\begin{equation}
 {\bf 248}_{E_8}\rar\left[\ ( {\bf 1}, {\bf 45})\right.+( {\bf 4}, {\bf 16})+(\overline{ {\bf 4}},\overline{ {\bf 16}})+( {\bf 6}, {\bf 10})+( {\bf 15}, {\bf 1})\left.\right]_{SU(4)\times SO(10)}
 \end{equation} 
These various representations further decompose under $SU(4)\times SU(5)\times U_2(1)$ as
\begin{eqnarray}
({\bf 1},{\bf 45})&\rar&({\bf 1},{\bf 1})_0+({\bf 1},{\bf 10})_4+({\bf 1},\overline{\bf 10})_{-4}+({\bf 1},{\bf 24})_0\label{(1,45)}\\
({\bf 4},{\bf 16})&\rar&({\bf 4},{\bf 1})_{-5}+({\bf 4},\overline{\bf 5})_3+({\bf 4},{\bf 10})_{-1}\label{(4,16)}\\
(\overline{\bf 4},\overline{\bf 16})&\rar&(\overline{\bf 4},{\bf 1})_{5}+(\overline{\bf 4},{\bf 5})_{-3}+(\overline{\bf 4},\overline{\bf 10})_{1}\quad\\
({\bf 6},{\bf 10})&\rar& ({\bf 6},{\bf 5})_2+({\bf 6},\overline{\bf 5})_{-2}\label{(6,10)}\\
({\bf 15},{\bf 1})&\rar& ({\bf 15},{\bf 1})_0\quad\label{(15,1)}
\end{eqnarray}
We now turn to the second group decomposition~\eqref{decomp2}. Here we have
\begin{equation}
{\bf  248}_{E_8}\rar\left[\ ({\bf 1},{\bf 24})\right.+({\bf 24},{\bf 1})+({\bf 10},\overline{{\bf 5}})+({\bf 5},{\bf 10})+(\overline{{\bf 10}},{\bf 5})+(\overline{{\bf 5}},\overline{{\bf 10}})\left.\right]_{SU(5)\times SU(5)}\; .
\end{equation} 
Further branching under $SU(5)\times SU(5)\times U_2'(1)$ leads to
\begin{eqnarray}
({\bf 1},{\bf 24})&\rar&({\bf 1},{\bf 24})_0\\
({\bf 24},{\bf 1})&\rar&({\bf 1},{\bf 1})_0+({\bf 4},{\bf 1})_{-5}+(\overline{\bf 4},{\bf 1})_5+({\bf 15},{\bf 1})_0\\
({\bf 10},\overline{\bf 5})&\rar&({\bf 4},\overline{\bf 5})_3+({\bf 6},\overline{\bf 5})_{-2}\\
({\bf 5},{\bf 10})&\rar&({\bf 1},{\bf 10})_4+({\bf 4},{\bf 10})_{-1}\\
(\overline{\bf 10},{\bf 5})&\rar&(\overline{\bf 4},{\bf 5})_{-3}+({\bf 6},{\bf 5})_2\\
(\overline{\bf 5},\overline{\bf 10})&\rar&({\bf 1},\overline{\bf 10})_{-4}+(\overline{\bf 4},\overline{\bf 10})_1
\end{eqnarray}
A comparison of these two decompositions shows that, in fact, $U_2(1)=U_2'(1)$. We will denote this symmetry by $U_X(1)$ with charge $X$ from hereon. The charge $Q_1$ of $U_1(1)$ is related to the weak hypercharge by $Q_1 = 3Y$. We also define the combination
\begin{equation}
\quad Q=\frac{1}{15}(2Q_1-3X)=\frac{1}{5}(2Y-X)\; , \label{Qdef}
\end{equation}
On the matter fields descending from the spinor of $SO(10)$, that is the mulitplets $({\bf 4},{\bf 1})_{-5}$, $({\bf 4},\bar{\bf 5})_3$, $({\bf 4},{\bf 10})_{-1}$ from Eq.~\eqref{(4,16)}, the charge $Q$ is identical to $B-L$. Further, for the multiplets descending from the vector of $SO(10)$, that is $({\bf 6},{\bf 5})_2$, $({\bf 6},\overline{\bf 5})_{-2}$ in Eq.~\eqref{(6,10)}, the $Q$ charges of the $SU(2)$ doublets vanish, so this provides the correct $B-L$ charge for the Higgs multiplets. We refer to all these fields for which $Q$ provides the standard value of $B-L$  as {\it regular}. For all other matter fields in Eqs.~\eqref{(1,45)}--\eqref{(15,1)} $Q$ is different from $B-L$ and we refer to these fields as {\it irregular}. In Table~\ref{Table:spectrum} we summarise the $SU(4)\times SU(5)\times U_X(1)$ multiplets and their associated bundles, whose first cohomology groups count the number of multiplets of each type, along with the charge $Q$. Multiplets are denoted by their standard name and irregular multiplets are indicated by a prime.
\begin{table}
\begin{center}
\begin{tabular}{|l|l|l|l|l|c|c|l|}\hline
{\fn \varstr{5pt}{3pt}$SU(4)\times$}
&{\fn $SU(4)\times$}&{\fn $SU(4)\times SU_c(3)\times$}&{\fn symbol}&{\fn name}&{\fn $Y$}&{\fn $Q$}&{\fn $H^1(X,\cdot)$}\\
{\fn \varstr{3pt}{3pt}$SO(10)$}&{\fn $SU(5)_{U_X(1)}$}&{\fn $SU_W(2)_{U_1(1),U_X(1)}$}&&&&&\\\hline\hline\hline
\varstr{5pt}{3pt}$({\bf 1},{\bf 45})$&$({\bf 1},{\bf 1})_0$&$({\bf 1},{\bf 1},{\bf 1})_{0,0}$&${N^c}'$&\fn irr.~RN neutrino&$0$&$0$&${\cal O}_X$\\\hline
\varstr{5pt}{3pt}&$({\bf 1},{\bf 10})_4$&$({\bf 1},{\bf 1},{\bf 1})_{6,4}$&${e^c}'$&\fn irr.~RH electron&$2$&$0$&$L$\\
\varstr{5pt}{3pt}&&$({\bf 1},\overline{\bf 3},{\bf 1})_{-4,4}$&${u^c}'$&\fn irr.~RH u quark&$-4/3$&$-4/3$&\\
\varstr{5pt}{3pt}&&$({\bf 1},{\bf 3},{\bf 2})_{1,4}$&$Q'$&\fn irr.~LH quarks&$1/3$&$-2/3$&\\\hline
\varstr{5pt}{3pt}&$({\bf 1},\overline{\bf 10})_{-4}$&$({\bf 1},{\bf 1},{\bf 1})_{-6,-4}$&${\tilde e}^{c\prime}$&\fn irr.~RH mirror electron&$-2$&$0$&$L^*$\\
\varstr{5pt}{3pt}&&$({\bf 1},{\bf 3},{\bf 1})_{4,-4}$&${\tilde u}^{c\prime}$&\fn irr.~RH mirror u quark&$4/3$&$4/3$&\\
\varstr{5pt}{3pt}&&$({\bf 1},\overline{\bf 3},{\bf 2})_{-1,-4}$&${\widetilde Q}'$&\fn irr.~LH mirror quarks&$-1/3$&$2/3$&\\\hline
\varstr{5pt}{3pt}&$({\bf 1},{\bf 24})_0$&&&\fn $SU(5)$ gauge bosons&&&${\cal O}_X$\\\hline\hline
\varstr{5pt}{3pt}$({\bf 4},{\bf 16})$&$({\bf 4},{\bf 1})_{-5}$&$({\bf 4},{\bf 1},{\bf 1})_{0,-5}$&$N^c$&\fn RH neutrino&$0$&$1$&$U\otimes L^*$\\\hline
\varstr{5pt}{3pt}&$({\bf 4},\overline{\bf 5})_3$&$({\bf 4},{\bf 1},{\bf 2})_{-3,3}$&$L $&\fn LH leptons &$-1$&$-1$&$U\otimes L$\\
\varstr{5pt}{3pt}&&& $H'$&\fn  irr.~d-Higgs &$-1$&$-1$&\\
\varstr{5pt}{3pt}&&$({\bf 4},\overline{\bf 3},{\bf 1})_{2,3}$&$d^c $&\fn RH d quark &$2/3$&$-1/3$&\\
\varstr{5pt}{3pt}&&& $T'$&\fn  irr.~d-Higgs triplet&$2/3$&$-1/3$&\\
\hline
\varstr{5pt}{3pt}&$({\bf 4},{\bf 10})_{-1}$&$({\bf 4},{\bf 1},{\bf 1})_{6,-1}$&$e^c$&\fn RH electron&$2$&$1$&$U$\\
\varstr{5pt}{3pt}&&$({\bf 4},\overline{\bf 3},{\bf 1})_{-4,-1}$&$u^c$&\fn RH u quark&$-4/3$&$-1/3$&\\
\varstr{5pt}{3pt}&&$({\bf 4},{\bf 3},{\bf 2})_{1,-1}$&$Q$&\fn LH quarks&$1/3$&$1/3$&\\\hline\hline
\varstr{5pt}{3pt}$(\overline{\bf 4},\overline{\bf 16})$&$(\overline{\bf 4},{\bf 1})_{5}$&$(\overline{\bf 4},{\bf 1},{\bf 1})_{0,5}$&${\widetilde N}^c$&\fn RH mirror neutrino&$0$&$-1$&$U^*\otimes L$\\\hline
\varstr{5pt}{3pt}&$(\overline{\bf 4},{\bf 5})_{-3}$&$(\overline{\bf 4},{\bf 1},{\bf 2})_{3,-3}$&$\widetilde{L} $&\fn LH mirror leptons &$1$&$1$&$U^*\otimes L^*$\\
\varstr{5pt}{3pt}&&&$\overline{H}'$ &\fn irr.~u-Higgs &$1$&$1$&\\ 
\varstr{5pt}{3pt}&&$(\overline{\bf 4},{\bf 3},{\bf 1})_{-2,-3}$&${\tilde d}^c$&\fn RH mirror d quark &$-2/3$&$1/3$&\\
\varstr{5pt}{3pt}&&&$ \overline{T}'$&\fn   irr.~u-Higgs triplet&$-2/3$&$1/3$&\\
\hline
\varstr{5pt}{3pt}&$(\overline{\bf 4},\overline{\bf 10})_{1}$&$(\overline{\bf 4},{\bf 1},{\bf 1})_{-6,1}$&${\tilde e}^c$&\fn RH mirror electron&$-2$&$-1$&$U^*$\\
\varstr{5pt}{3pt}&&$(\overline{\bf 4},{\bf 3},{\bf 1})_{4,1}$&${\tilde u}^c$&\fn RH mirror u quark&$4/3$&$1/3$&\\
\varstr{5pt}{3pt}&&$(\overline{\bf 4},\overline{\bf 3},{\bf 2})_{-1,1}$&$\widetilde{Q}$&\fn LH mirror quarks&$-1/3$&$-1/3$&\\\hline\hline
\varstr{5pt}{3pt}$({\bf 6},{\bf 10})$&$({\bf 6},{\bf 5})_2$&$({\bf 6},{\bf 1},{\bf 2})_{3,2}$&$\overline{H} $&\fn u Higgs &$1$&$0$&$\wedge^2U^*$\\
\varstr{5pt}{3pt}&&&$\widetilde{L}'$ &\fn irr.~LH mirror leptons &$1$&$0$&\\ 
\varstr{5pt}{3pt}&&$({\bf 6},{\bf 3},{\bf 1})_{-2,2}$&$\overline{T} $&\fn Higgs triplet &$-2/3$&$-2/3$&\\
\varstr{5pt}{3pt}&&&$\tilde{d}^{c'}$&\fn  irr.~RH mirror d quark &$-2/3$&$-2/3$&\\
\hline
\varstr{5pt}{3pt}&$({\bf 6},\overline{\bf 5})_{-2}$&$({\bf 6},{\bf 1},{\bf 2})_{-3,-2}$&$H $&\fn d Higgs &$-1$&$0$&$\wedge^2U$\\
\varstr{5pt}{3pt}&&&$L'$&\fn  irr.~LH leptons &$-1$&$0$&\\ 
\varstr{5pt}{3pt}&&$({\bf 6},\overline{\bf 3},{\bf 1})_{2,-2}$&$T $&\fn Higgs triplet &$2/3$&$2/3$&\\
\varstr{5pt}{3pt}&&&$d^{c'}$ &\fn irr.~RH d quark &$2/3$&$2/3$&\\ 
\hline
\end{tabular}
\caption{\it Particle content resulting from compactifications with $S(U(4)\times U(1))$-bundles $V=U\oplus L$. The particles labelled with a prime and referred to as ``irregular" have the same quantum numbers under the SM group as the corresponding un-primed particles but, unlike for those, their $Q$ charge is different from  $B-L$.}
\label{Table:spectrum}
\end{center}
\end{table}

It is also useful to collect expressions for the chiral asymmetries of the various multiplets. We denote these by $N(R)$ by which we mean the number of multiplets in the representation $R$ minus the number of multiplets in the representation $\bar{R}$ of $SU(4)\times SU(5)\times U_X(1)$. Using Table~\ref{Table:spectrum}, the results from Appendix A and the fact that ${\rm c}_1(U)=-{\rm c}_1(L)$, we find for the ${\bf 10}_{SU(5)}$ multiplets
\begin{eqnarray}
 N\left(({\bf 4},{\bf 10})_{-1}\right)&=&\ind(U^*)=-{\rm ch}_3 (U) + \frac{1}{12} {\rm c}_1 (L ) {\rm c}_2 (TX)\label{indU}\\
 N\left(({\bf 1},{\bf 10})_4\right)&=&\ind(L^*)=-\frac{1}{12} \left( 2 {\rm c}_1 (L)^3 + {\rm c}_1 (L) {\rm c}_2 (TX)\right)\; , \label{indL}
\end{eqnarray}
where the first line counts the chiral asymmetry for the regular ${\bf 10}$ multiplets and the second one the number of irregular ones. Similarly, with ${\rm c}_1(U)=-{\rm c}_1(L)$ and Appendix A, we have for the $\overline{\bf 5}$ multiplets
\begin{eqnarray}
N\left(({\bf 4},\bar{\bf 5})_{3}\right)&=&\ind(U^*\otimes L^*)=-\ind(U)-4\ind(L)+\frac{1}{2}{\rm c}_1(L)^3-{\rm c}_1(L)\ch_2(U)\label{indUL}\\
N\left(({\bf 6},\bar{\bf 5})_{-2}\right)&=&\ind(\wedge^2U^*)={\rm c}_1(L)\ch_2(U)-\frac{1}{4}{\rm c}_2(TX){\rm c}_1(L)
=3\ind(L)-c_1(L)c_2(U)\; ,\label{indW2U}
\end{eqnarray}
where the first line counts the asymmetry in regular $\bar{\bf 5}$ multiplets and the second line the asymmetry in irregular ones. Finally, we have the asymmetry of regular right-handed neutrinos given by~\footnote{From Table~\ref{Table:spectrum}, the irregular right-handed neutrinos are completely uncharged and will not play any role for our discussion.}
\begin{equation}
 N\left(({\bf 4},{\bf 1})_{-5}\right)=\ind(U^*\otimes L)=-\ind(U)+4\ind(L)+\frac{1}{2}{\rm c}_1(L)^3+{\rm c}_1(L)\ch_2(U)\; .\label{indRHN}
\end{equation} 
Rank five bundles $V$ with ${\rm c}_1(V)=0$ satisfy in general $\ind(V)=\ind(\wedge^2V)$. For split bundles $V=U\oplus L$ with $\wedge^2V=\wedge^2U\oplus (U\otimes L)$ this implies that the total asymmetry of (regular and irregular) ${\bf 10}$ and $\overline{\bf 5}$ multiplets is the same since
\begin{equation} 
 N\left({\bf 10}\right)=\ind(U^*)+\ind(L^*)=\ind (V^*)=\ind(\wedge^2V^*)=\ind(\wedge^2U^*)+\ind(U^*\otimes L^*)=N\left(\bar{\bf 5}\right)\; .
\end{equation} 
Hence, such models always lead to complete chiral families in ${\bf 10}\oplus\bar{\bf 5}$. The chiral asymmetry~\eqref{indRHN} for the regular right-handed neutrinos, however, is in general independent and not linked to the number of families. 

The low-energy gauge group of these model is $SU(5)\times U_X(1)$, before Wilson-line breaking of the GUT symmetry, and one generally expects the $U_X(1)$ symmetry to be Green-Schwarz anomalous. The anomalies can be computed in terms of the above chiral asymmetries. For the mixed $U_X(1)-SU(5)^2$ anomaly $A_1$ and the cubic $U_X(1)^3$ anomaly $A_3$ one finds, respectively,
\begin{eqnarray}
 A_1&=&-\sum_R X(R)\,c(R)\, N(R)\nonumber\\
       &=&12\ind(L)+3\ind(U\otimes L)-3\ind(U)-2\ind(\wedge^2U)\label{A1}\\[2mm]
 A_3&=&-\sum_R X(R)^3{\rm dim}(R)\, N(R)\nonumber\\
       &=&640\ind(L)-125\ind(U\otimes L^*)+135\ind(U\otimes L)-10\ind(U)-40\ind(\wedge^2 U)\label{A3}\; ,
\end{eqnarray}
where we have used the values $c(5_{SU(5)})=c(\bar{5}_{SU(5)})=1$ and $c(10_{SU(5)})=c(\bar{10}_{SU(5)})=3$ for the group-theoretical indices of $SU(5)$. The presence of the additional $U_X(1)$ factor in the low energy gauge group could, in principle, lead to unobserved long range interactions. However, this is not the case here. The gauge boson associated to the $U_X(1)$ symmetry is always massive~\cite{Witten:1985bz, Anderson:2009nt} with a mass given by
\be 
m^2_{U(1)} \sim G_{i j}\, {\rm c}_{1}^{ i}({L})\, {\rm c}_{1}^{j}({L}) 
\label{eq:u1mass}
\ee
where $G_{i j}$ is the K\"ahler moduli space metric. Since the K\"ahler moduli space metric is positive definite, it follows that $m^2_{U(1)}>0$, with a typical magnitude of order of the string scale, as long as the line bundle $L$ is non-trivial, which we assume. 

What about the constraints we need to impose in order to obtain a realistic particle spectrum? First, in order to have three chiral families of ${\bf 10}$ multiplets we need $\ind(V)=\ind(U)+\ind(L)\stackrel{!}{=}-3|\Gamma|$ where $|\Gamma|$ is the order of a freely-acting discrete symmetry $\Gamma$ which the model is to be divided by. From the above argument this automatically guarantees the same chiral asymmetry for the $\bar{\bf 5}$ multiplets. In addition, we should require the absence of $\overline{\bf 10}$ mirror families, that is, $h^1(V^*)=h^1(U^*)+h^1(L^*)\stackrel{!}{=}0$. We do not need to impose the analogous constraint on the ${\bf 5}$--$\bar{\bf 5}$ sector since a pair of Higgs doubles needs to emerge from this sector. Instead, we require the presence of at least one vector-like ${\bf 5}$--$\bar{\bf 5}$ pair, which means that $h^1(\wedge^2V^*)=h^1(\wedge^2U^*)+h^1(U^*\otimes L^*)\stackrel{!}{>}0$. So, to summarize, for a realistic spectrum we require
\begin{align}
 \mbox{three chiral families: }\quad&\ind(V)=\ind(U)+\ind(L)\stackrel{!}{=}-3|\Gamma|\label{phys1}\\
 \mbox{absence of }\overline{\bf 10}\mbox{ mirror families: }\quad&h^1(V^*)=h^1(U^*)+h^1(L^*)\stackrel{!}{=}0\\
 \mbox{presence of }{\bf 5}-\bar{\bf 5}\mbox{ pair to account for Higgs: }\quad&h^1(\wedge^2V^*)=h^1(\wedge^2U^*)+h^1(U^*\otimes L^*)\stackrel{!}{>}0\label{phys3}
\end{align} 
Models satisfying these constraints will lead to a low-energy theory with three families and, subject to successfully projecting out the Higgs triplets by the Wilson line while keeping a pair of doublets, with the correct spectrum of Higgs fields. However, in general these models can still contain regular fields, where the charge $Q$ is the standard $B-L$ charge as well as irregular fields, for which $Q$ takes a value different from $B-L$. In fact, in the above constraints~\eqref{phys1}--\eqref{phys3}, the first term in the middle corresponds to the regular contribution while the second one counts the irregular multiplets. While all models have an additional $U(1)$ symmetry with charge $Q$, this symmetry only coincides with $B-L$ for models which contain regular multiplets only. We will refer to such models with regular multiplets only as {\it regular models} and we will focus on such models in the following.
\section{Regular models}\label{sec:regmodels}
In addition to the conditions~\eqref{phys1}--\eqref{phys3} for models with a realistic spectrum, for regular models we need to impose that
\begin{equation}
 \ind(L)\stackrel{!}{=}\ind(\wedge^2 U)\stackrel{!}{=}0\; , \label{regfam}
\end{equation} 
so that all ${\bf 10}\oplus\bar{\bf 5}$ families are regular. In this case, all ${\bf 10}$ multiplets originate from $H^1(X,U)$, while all $\bar{\bf 5}$ multiplets originate from $H^1(X,U\otimes L)$. In addition, in order to have regular Higgs doublets, we have to ensure that they originate from $H^1(X,\wedge^2U)$ and $H^1(X,\wedge^2U^*)$ only, for which we have to demand in addition that
\begin{equation}
 h^1(U^*\otimes L^*)\stackrel{!}{=}0\; . \label{reghiggs}
\end{equation} 
Imposing the regularity conditions~\eqref{regfam} on Eqs.~\eqref{indL} and \eqref{indW2U} we obtain 
\begin{align}
0& = {\rm Ind} (L)= \frac{1}{12} \left(2\, {\rm c}_1 (L)^3 +  {\rm c}_1 (L)\,  {\rm c}_2 (TX)\right)\\ 
0&= {\rm Ind} (\wedge^2 U)= -3\, {\rm Ind} (L) + {\rm c}_1 (L)\, {\rm c}_2 (U)= {\rm c}_1 (L)\, {\rm c}_2 (U)\,. 
\label{eq:indices1}
\end{align}
Using these relations together with Eqs.~\eqref{indU}, \eqref{indUL}, \eqref{indRHN} and the results from Appendix A we find that the chiral asymmetries for regular models simplify to
\bea
N({\bf 10})&=&{\rm Ind} (U^*)=- \frac{1}{2}{\rm c}_3 (U) \\
N(\bar{\bf 5})&=&{\rm Ind} (U^* \otimes L^*) = - \frac{1}{2}{\rm c}_3 (U)\\
N({\bf 1})&=&{\rm Ind} (U^* \otimes L) = -\frac{1}{2} {\rm c}_3 (U)+ {\rm c}_1 (L)^3\; .
\eea
So we have an equal chiral asymmetry for ${\bf 10}$ and $\bar{\bf 5}$ multiplets, as expected, while the asymmetry for right-handed neutrinos remains independent. With these results and \eqref{regfam}, the expressions~\eqref{A1} and \eqref{A3} for the $U_X(1)$ anomalies simplify to
\begin{eqnarray}
 A_1&=&3\ind(U\otimes L)-3\ind(U)=0\label{A1reg}\\
 A_3&=&125(\ind(U)-\ind(U\otimes L^*))=125\,{\rm c}_1(L)^3\; ,\label{A3reg}
\end{eqnarray}
that is, the mixed $U_X(1)$--$SU(5)^2$ anomaly always vanished for regular models, as is expected for a $B-L$ symmetry, while the cubic $U_X(1)^3$ anomaly is controlled by the difference between the chiral asymmetries of families and right-handed neutrinos. In particular, if there are as many chiral right-handed neutrinos as chiral families the cubic anomaly vanishes as well.
\section{The $SU(5)\times U(1)$ model}
\label{sec:GUTmodel}
In~Ref.\cite{Buchbinder:2013dna} we have constructed an example of a regular model in the above sense, at the GUT level. 
The purpose of this section is to review this construction as a preparation for the discussion of the associated standard model in the next section.

\subsection{The manifold}
We begin by reviewing the construction of the model presented in Ref.~\cite{Buchbinder:2013dna}. The compactification space $X$  is chosen to be the tetra-quadric manifold, that is, a smooth Calabi-Yau hypersurface embedded in a product of four complex projective spaces, ${\cal A}= \left({\mathbb P}^1\right)^{\times 4}$ defined as the zero locus of a polynomial that is quadratic in the coordinates of each projective space. This manifold $X$ has the following Hodge numbers.
\begin{equation}
h^{1, 1} (X)=4\,, \quad h^{1, 2} (X)=68\,, 
\label{2.1}
\end{equation}
The pullbacks, $J_i$, of the four canonically normalized $\mathbb{P}^1$ K\"ahler forms to $X$ provide a basis of the second cohomology of $X$. In terms of this basis, the triple intersection numbers are
\be
d_{ijk} = \int_X J_i\wedge J_j\wedge J_k = \begin{cases} 2 & \mbox{ if } i\neq j, j\neq k \\ 0 &\mbox{ otherwise } \end{cases}
\label{dijk}
\ee
and the second Chern class is given by
\be
{\rm c}_2 (TX)= (24, 24, 24, 24)\,. 
\label{2.2.1}
\ee
The K\"ahler forms on $X$ are parametrized as $J=t^iJ_i$ where $t^i$ are the K\"ahler moduli. The K\"ahler cone is the portion of $t^i$ space where all $t^i\geq 0$. It is useful to also introduce the dual K\"ahler moduli $s_i = d_{i j k}t_j t_k$. In terms of these dual variables, the K\"ahler cone can be expressed as
\be 
C_{K}=\{ \vec{s} \in {\mathbb R}^4 | \ {\bf n}_i \cdot {\bf s} \geq 0, \ {\bf e}_i \cdot {\bf s}_i \geq 0\}\,, 
\label{2.3}
\ee
where ${\bf e}_i$ are the standard unit vectors in ${\mathbb R}^4$, ${\bf n}_i = {\bf n}- {\bf e}_i $ and $ {\bf n}= \frac{1}{2} (1, 1, 1, 1)$.

\subsection{The bundle}
We construct the vector bundle $V$ on $X$ as the direct sum
\be
 V=U\oplus L
\ee
of the line bundle
\be
L={\cal O}_X (1, 1, -1, -1)\label{Ldef}
\ee
and the rank four bundle $U$, defined by the monad sequence
\be 
0 \longrightarrow U \longrightarrow B \stackrel{f}{\longrightarrow} C \longrightarrow 0\quad\mbox{where}\quad
B=\bigoplus_{b=1}^{r_B}{\cal O}_X({\boldsymbol \beta}_b)\;,\quad C=\bigoplus_{c=1}^{r_C}{\cal O}_X({\boldsymbol \gamma}_c)\; .
\label{monaddef}
\ee
Here, $B$ and $C$ are line bundle sums of ranks $r_B=6$ and $r_C=2$, respectively, which are explicitly given by
\begin{equation}
\begin{aligned} 
B\,  = &\ {\cal O}_X (-1, 0, 0, 1) \oplus {\cal O}_X (-1, -1, 2, 2)\oplus {\cal O}_X (-1, -1, 2, 2)\\
& \oplus{\cal O}_X (0, 1, -1, 0) \oplus {\cal O}_X (1, 1, 0, 0)\oplus {\cal O}_X (1, 1, 0, 0)\,, \\
C\, = &\ {\cal O}_X (-1, 1, 2, 2) \oplus {\cal O}_X (1, 1, 0, 2)\,. 
\label{BCdef}
\end{aligned}
\end{equation}
The most generic map consistent with the above choice for the line bundle sums $B$ and $C$ is given by 
\begin{equation}
 f = \left(\begin{array}{lllllll}f_{(0,1,2,1)}&f_{(0,2,0,0)}&f_{(0,2,0,0)}'&0&0&0\\
 f_{(2,1,0,1)}&0&0&f_{(1,0,1,2)}&f_{(0,0,0,2)}&f_{(0,0,0,2)}'\end{array}\right)\,,
\label{eq:monadmap}
\end{equation} 
where the subscripts indicate multi-degrees of polynomials. For example, $f_{(0,1,2,1)}$ is of degree 
0 in the first~${\mathbb P}^1$, degree 1 in the second  ${\mathbb P}^1$, and so on.

In general, for a monad bundle $U$ as above, the total Chern character satisfies $\ch(B)=\ch(U)+\ch(C)$. Combining this with the equations in Appendix A it can be shown that the Chern characters of $U$ are explicitly given by
\bea
{\rm rk} (U)&=& r_B - r_C\nonumber \\
\ch_1^i (U) &=& \sum_{b=1}^{r_B} \beta_b^i - \sum_{c=1}^{r_C} \gamma_c^i\nonumber \\
\ch_{2 i}(U) &=&\frac{1}{2} d_{i j k} \left ( \sum_{b=1}^{r_B} \beta_b^j \beta_b^k -\sum_{c=1}^{r_C} \gamma_{c}^j \gamma_c^k\right)\nonumber \\
\ch_{3}(U) &=&\frac{1}{6} d_{i jk} \left ( \sum_{b=1}^{r_B} \beta_b^i \beta_{b}^j \beta_b^k - \sum_{c=1}^{r_C} \gamma_c^i \gamma_c^j 
 \gamma_c^k \right)\,. 
 \label{2.9.02}
 \eea
Inserting the explicit line bundles~\eqref{BCdef} of our model into the above formulae and using Eq.~\eqref{A1} we find the Chern classes of $V$ are given by
\be
 {\rm c}_1(V)=0\;,\quad {\rm c}_2(V)=(24,8,20,12)\;,\quad {\rm c}_3(V)=-24\; .
\ee
Note, since
\be 
\quad {\rm c}_2 (V)= (24, 8, 20, 12) \leq {\rm c}_2(TX)=(24,24,24,24) \; ,
\label{2.9.1}
\ee
the anomaly cancellation condition can indeed be satisfied.

\subsection{Slope stability}
The bundle $V$ is supersymmetric if it is poly-stable and has slope zero. Poly-stability means that each non-decomposable part of $V$ is stable and has the same slope as the total bundle $V$, that is, zero in the present case. In general, the slope of a bundle (or sheaf) ${\cal F}$ is the K\"ahler moduli dependent quantity 
\begin{equation}
\mu({\cal F}) = \frac{1}{\text{rk} ({\cal F})}\, d_{i j k}\,{\rm c}_1^i({\cal F})\, t^j\, t^k=\frac{1}{\text{rk} ({\cal F})}\,{\bf s}\cdot {\rm c}_1({\cal F})\; .
\end{equation}
Since ${\rm c}_1(V)=0$, the slope of the full bundle $V$ vanishes automatically. Vanishing of the slope of $L$ imposes the condition
\be 
\mu (L) = {\bf s}\cdot (1, 1, -1, -1) \stackrel{!}{=}0 \,. 
\label{2.10}
\ee
on the K\"ahler parameters of the theory, that is, we should restrict the model to the so-defined co-dimension one locus in K\"ahler moduli space. Since ${\rm c}_1(U)=-{\rm c}_1(L)$, the vanishing of the slope of $U$ is then automatic. Line bundles are always stable so nothing further needs to be checked for $L$. The verify stability of $U$ we have to check that all sub-sheafs ${\cal F}\subset U$ with $0<{\rm rk}({\cal F})<{\rm rk}(U)$ satisfy $\mu({\cal F})<\mu(U)=0$, so all proper sub-sheafs of $U$ need to have a strictly negative slope. This was studied in detail in Ref.~\cite{Buchbinder:2013dna} for a generic monad map of the type \eqref{eq:monadmap} using techniques described in Refs.~\cite{Anderson:2008ex,Anderson:2008uw}. Within the hyperplane defined by Eq.~\eqref{2.10}, the region of stability for $U$, and thus the region of poly-stability for $V$ is given by the cone
\be 
C_V=\{ {\bf s}\in C_K\ |\ {\bf s}\cdot (1, 1, -1, -1) =0, \ {\bf s} \cdot (-1, 1, 0, 0) \geq 0, \ {\bf s} \cdot (1, 1, -2, 0) \geq 0\}\, \label{CU}
\ee
In conclusion, the bundle $V$ is supersymmetric in the dimension three cone defined by Eq.~\eqref{CU}.

\subsection{The GUT spectrum}
The GUT spectrum of the above model is derived from the relevant cohomology groups in Table~\ref{Table:spectrum}. They were computed in detail in Ref.~\cite{Buchbinder:2013dna} and are explicitly given by~\footnote{All results in Eq.~\eqref{2.17} are obtained from explicit cohomology calculations except for $h^{1} (X, U \otimes U^*)$ which is inferred from arguments based on the low-energy effective theory, as explained later.}
\begin{eqnarray}
h^{\bullet} (X, U) =(0, 12, 0, 0)\,, \quad & h^{\bullet} (X, L) =(0, 0, 0, 0)\,, \quad & h^{\bullet} (X, U \otimes L) =(0, 12, 0, 0)\,, \nonumber\\
h^{\bullet} (X, \wedge^2 U) =(0, 3, 3, 0)\,, \quad &h^{\bullet} (X, U \otimes L^*) =(0, 12, 0, 0)\,, \quad &
h^{1} (X, U \otimes U^*) =45\,.
\label{2.17}
\end{eqnarray}
First, we note that these cohomologies satisfy the conditions~\eqref{phys1}--\eqref{phys3} for a physical spectrum provided the order of the freely-acting symmetry group $\Gamma$ is $|\Gamma|=4$. Further, the regularity conditions~\eqref{regfam}, \eqref{reghiggs} are satisfied so we are indeed dealing with a regular model. Hence, comparing the above cohomologies with Table~\ref{Table:spectrum}, we have a GUT theory with gauge group $SU(5)\times U_X(1)$ and spectrum
\begin{equation}
 12\,{\bf 10}_{-1}\,,\; 12\,\bar{\bf 5}_3\,,\; 3\,\bar{\bf 5}_{-2}\,,\; 3\,{\bf 5}_2\,,\; 12\,{\bf 1}_{-5}\,,\; 45\,{\bf 1}_0\; .\label{NonAbSU(5)spec}
\end{equation}
This amounts to $12$ families in ${\bf 10}\oplus\bar{\bf 5}$, three vector-like pairs ${\bf 5}\oplus\bar{\bf 5}$ to account for the Higgs doublets and $12$ right-handed neutrinos. Since the model is regular, the $U_X(1)$ symmetry combines with hypercharge to the standard $B-L$ symmetry, as in Eq.~\eqref{Qdef}. As discussed earlier, the $U_X(1)$ gauge boson is massive and, therefore, not of phenomenological concern. A special feature of our model is that the number of right-handed neutrinos equals the number of families, a property which does not have to be satisfied for regular models in general. From Eq.~\eqref{A3reg}, this means that the cubic $U_X(1)^3$ anomaly vanishes and since the mixed $U_X(1)-SU(5)^2$ anomaly vanishes for all regular models (see Eq.~\eqref{A1reg}) the additional $U_X(1)$ symmetry is entirely anomaly-free for this model. Another way to verify the vanishing of the cubic anomaly, using Eq.~\eqref{A3reg}, is to check that $c_1(L)^3=d_{ijk}{\rm c}_1^i(L){\rm c}_1^j(L){\rm c}_1^k(L)=0$. This is indeed the case, in view of our specific choice~\eqref{Ldef} of $L$ and the triple intersection numbers~\eqref{dijk}.
Note that, even thought the $U_X(1)$ symmetry is non-anomalous in our specific model, the associated gauge boson is still super-massive, as was pointed out above Eq.~\eqref{eq:u1mass}.

\subsection{The Abelian locus}
At a generic point in the poly-stable region~\eqref{CU} of our bundle $V$ the structure group is $S(U(4)\times U(1))$, however, this can split further at particular sub-loci. Indeed, it was shown in Ref.~\cite{Buchbinder:2013dna} that for a monad bundle map~\eqref{eq:monadmap} satisfying 
\be 
f_{(0, 1, 2, 1)}=f_{(2, 1, 0, 1)}= f_{(1, 0, 1, 2)}=0\,. \label{fcond}
\ee
the bundle $U$ splits up into a line bundle sum $U_0$. More specifically, the total bundle $V$ splits into a sum of five line bundles
\begin{equation}
 V_0=U_0\oplus L\;,\quad U_0= \bigoplus_{a=1}^4L_a\;,\quad L=L_5 \label{V0def}
\end{equation}
which are explicitly given by
\begin{equation}\label{Ladefs}
\begin{array}{lllllllllll}
 L_1&=&\cO_X(-1,0,0,1)&\quad&L_2&=&\cO_X(-1,-3,2,2)&\quad&L_3&=&\cO_X(0,1,-1,0)\\
 L_4&=&\cO_X(1,1,0,-2)&\quad&L_5&=&{\cal O}_X (1, 1, -1, -1)\; .&&&&
\end{array}
\end{equation} 
At the same time the bundle structure group reduces from $S(U(4)\times U(1)$ to $S(U(1)^5)$. In addition to the restrictions in bundle moduli space represented by~\eqref{fcond} this also requires a further restriction in K\"ahler moduli space since all line bundles in $V_0$ need to have vanishing slope. Since there exist two linear relations between the first Chern classes of the five line bundles, this split locus is described by three conditions which can be written as 
\be
 {\bf s}\cdot (1, 1, -1, -1) =0\,, \quad {\bf s} \cdot (-1, 1, 0, 0) = 0\,, \quad {\bf s} \cdot (1, 1, -2, 0) = 0\,. 
 \label{eq:abelianlocus}
 \ee
These conditions are solved for $s_1= s_2 =s_3 =s_4$ or, equivalently, $t_1=t_2=t_3=t_4$, so along the diagonal in K\"ahler moduli space. Note that this diagonal, which we refer to as the Abelian locus, is contained in and lies on the boundary of~\eqref{CU}, the space where the bundle $V$ is supersymmetric. When the structure group splits to $S(U(1)^5)$ on this locus the low-energy gauge group enhances from $SU(5)\times U_X(1)$ to $SU(5)\times S(U(1)^5)$. As we will see, this symmetry enhancement has important consequences for the model, even away from the Abelian locus. 

The line bundle cohomologies relevant to computing the spectrum are
\begin{equation}\label{lbcoh}
\begin{array}{rrlrrrlrrrl}
 h^1(X,L_2)&=&8&\quad& h^1(X,L_4)&=&4&\quad& h^1(X,L_2\otimes L_4)&=&3\\
 h^1(X,L_2^*\otimes L_4^*)&=&3&\quad&h^1(X,L_2\otimes L_5)&=&4&\quad&h^1(X,L_4\otimes L_5)&=&8\\
 h^1(L_2\otimes L_1^*)&=&12&\quad&h^1(L_4\otimes L_1^*)&=&12&\quad& h^1(L_2\otimes L_3^*)&=&20\\
 h^1(L_4\otimes L_3^*)&=&4&\quad&h^1(L_2\otimes L_5^*)&=&12\; .&&
 \end{array}
\end{equation} 
Here, first cohomologies of single line bundles $L_a$ lead to ${\bf 10}_a$ multiplets of $SU(5)$, where the subscript indicates that this multiplet has charge one under the $a^{\rm th}$ $U(1)$ symmetry and is uncharged under all others. Cohomologies $H^1(X,L_a\otimes L_b)$ ($H^1(X,L_a^*\otimes L_b^*)$) lead to multiplets $\bar{\bf 5}_{a,b}$ (${\bf 5}_{a,b}$) which carry charge $1$ ($-1$) under the $a^{\rm th}$ and $b^{\rm th}$ $U(1)$ and are uncharged under all others. Finally, $H^1(X,L_a\otimes L_b^*)$ leads to singlets ${\bf 1}_{a,b}$ with charge $1$ under the $a^{\rm th}$ $U(1)$  and charge $-1$ under the $b^{\rm th}$ $U(1)$. With this identifications, the above cohomologies imply the following spectrum
\begin{equation}
 8\,{\bf 10}_2\,,\; 4\,{\bf 10}_4\,,\; 8\bar{\bf 5}_{4,5}\,,\; 4\bar{\bf 5}_{2,5}\,,\; 3\,\bar{\bf 5}_{2,4}\,,\; 3\,{\bf 5}_{2,4}\,,\;
 12\,{\bf 1}_{2,1}\,,\; 12\,{\bf 1}_{4,1}\,,\; 20\,{\bf 1}_{2,3}\,,\; 12\,{\bf 1}_{2,5}\,,\;4\,{\bf 1}_{4,3} \; .\label{AbSU(5)spec}
\end{equation}  
Note that the $SU(5)$ charged part of this spectrum is identical to the one for the non-Abelian bundle given in Eq.~\eqref{NonAbSU(5)spec}, that is, it consists of $12$ families in ${\bf 10}\oplus\bar{\bf 5}$ and three vector-like pairs ${\bf 5}\oplus\bar{\bf 5}$. It is not surprising that the number of families remains unchanged as we deform to a non-Abelian bundle since the chiral part of the spectrum is protected by an index. The fact that the three vector-like ${\bf 5}\oplus\bar{\bf 5}$ pairs remain massless as well is non-trivial and one of the appealing features of the model. This means we have a chance of obtaining Higgs doublets and, hence, a full standard model spectrum even away from the Abelian locus. In the following we will show that this can indeed be made to work. 

A final remark concerns the number of singlet fields. At the Abelian locus with structure group $S(U(1)^5)$ the spectrum~\eqref{AbSU(5)spec} contains a total of $60$ singlets. Moving away from this locus to a non-Abelian bundle with structure group $S(U(4)\times U(1))$ implies the Higgsing of three $U(1)$ symmetries in the low-energy theory. Hence, three of the $60$ bundle moduli are used to form the required massive gauge supermultiplets and we expect $57$ remaining bundle moduli. This is indeed the total number of moduli in the spectrum~\eqref{NonAbSU(5)spec}.
 
\section{The standard model with $B-L$ symmetry}\label{sec:SM}
Now we construct the standard model associated to the GUT described in the previous section. This involves taking the quotient of the GUT model with a suitable freely-acting symmetry and the inclusion of a Wilson line.

\subsection{Discrete symmetry and equivariant line bundles}
As a preparation we first introduce the relevant freely-acting symmetry of the tetra-quadric and discuss the equivariance properties of line bundles.

As we have already mentioned, with $12$ families present in the GUT theory, we require a freely-acting symmetry $\Gamma$ of order $|\Gamma|=4$. Luckily, a suitable freely-acting symmetry, $\Gamma=\mathbb{Z}_2\times\mathbb{Z}_2$, is available on the tetra-quadric~\cite{Candelas:2008wb,Braun:2010vc,Candelas:2010ve}. Its two generators act as
\begin{equation}
 \left( \begin{array}{cc}
1 &~~0 \\
0 & -1  \end{array}\right)\,, \quad 
 \left( \begin{array}{cc}
0 & ~~1 \\
1 & ~~0  \end{array}\right)\,. 
\label{eq:G-action}
\end{equation}
simultaneously on the homogenoues coordinates of each $\mathbb{P}^1$ ambient space. Our standard model will be based on the quotient Calabi-Yau manifold $\tilde{X}=X/\Gamma$ with Hodge numbers $h^{1,1}(\tilde{X})=4$ and $h^{2,1}(\tilde{X})=20$ and a non-trivial first fundamental group equal to $\Gamma=\mathbb{Z}_2\times\mathbb{Z}_2$.

For a well-defined model downstairs, we also have to ensure that the upstairs bundle $V$ on $X$ descends to a bundle $\tilde{V}$ on the quotient $\tilde{X}$. This is equivalent to saying that $V$ has a $\Gamma$--equivariant structure. Since line bundles are our basic building blocks our first step is to discuss the existence of equivariant structures for line bundles on the tetra-quadric. This is relatively easily done for line bundles which are globally generated by their sections. From the equivariant globally generated line bundles all equivariant line bundles can then be generated by applying conjugation and tensor products. This leads to the following generating list
\beq
\{\cO_X,\cO_X(2,0,0,0),\cO_X(1,1,0,0)\text{ and permutations thereof}\}\;,
\eeq 
for line bundles with equivariant structures under $\Gamma=\mathbb{Z}_2\times\mathbb{Z}_2$ with generators~\eqref{eq:G-action} on the tetra-quadric. In other words, all line bundles with such an equivariant structure can be obtained by taking arbitrary tensor products and conjugations of line bundles in the above list. We note that all line bundles used in our construction can be obtained in this way and, therefore, have an equivariant structure. This includes the line bundles in $B$, $C$, Eq.~\eqref{BCdef}, which were used to define the monad bundle $U$, the line bundle $L$ in Eq.~\eqref{Ldef} and the line bundles in $V_0$, Eq.~\eqref{V0def}, the bundle at the Abelian locus. The equivariant structure on a line bundle is not unique but can be multiplied by a one-dimensional representation of the discrete group. Hence, we can characterise the equivariant structure of a line bundle by an irreducible $\mathbb{Z}_2\times\mathbb{Z}_2$ representation. We denote these representations by pairs $(p,q)$ of charges, where $p,q=0,1$, so explicitly we have the four irreducible representations $(0,0),\, (0,1),\,(1,0),\,(1,1)$. We also denote the regular representation by
\be
 {\cal R}=(0,0)\oplus (0,1)\oplus(1,0)\oplus(1,1)\; .
\ee 
  
\subsection{Standard model at the Abelian locus}
As a warm-up it is useful to discuss the downstairs model at the Abelian locus first, before we move on to the general case. As a reminder, the vector bundle at the Abelian locus is a sum of five line bundles
\begin{equation}
 V_0=U_0\oplus L\;,\quad U_0= \bigoplus_{a=1}^4L_a\;,\quad L=L_5\; ,
\end{equation} 
where, from Eqs.~\eqref{V0def}, \eqref{Ladefs} and \eqref{Ldef}, the line bundles are explicitly given by
\begin{equation}\label{lbequiv}
\begin{array}{lllllllllll}
 L_1&=&\cO_X(-1,0,0,1)_{(0,0)}&\quad&L_2&=&\cO_X(-1,-3,2,2)_{(0,0)}&\quad&L_3&=&\cO_X(0,1,-1,0)_{(0,0)}\\
 L_4&=&\cO_X(1,1,0,-2)_{(0,0)}&\quad&L_5&=&{\cal O}_X (1, 1, -1, -1)_{(0,0)}\; .&&&&
\end{array}
\end{equation}  
Here the subscripts denote the $\mathbb{Z}_2\times\mathbb{Z}_2$ representations which specify the equivariant structure we have assigned to each line bundle. Our choice of the trivial representation for all line bundles will indeed turn out to be suitable. The dimensions of the relevant line bundle cohomologies have already been given in Eq.~\eqref{lbcoh}. Here, we need the decomposition of these cohomologies into  $\mathbb{Z}_2\times\mathbb{Z}_2$ representations. They can be found in the database~\cite{Linebundles} and are given by
\begin{equation}\label{lbcohgrad}
\begin{array}{rrlrrrlrrrl}
 H^1(X,L_2)&=&2\,{\cal R}&\quad& H^1(X,L_4)&=&{\cal R}&\quad& H^1(X,L_2\otimes L_4)&=& \tilde{\cal R}\\
 H^1(X,L_2^*\otimes L_4^*)&=& \tilde{\cal R}&\quad&H^1(X,L_2\otimes L_5)&=&{\cal R}&\quad&H^1(X,L_4\otimes L_5)&=&2\,{\cal R}\\
 H^1(L_2\otimes L_1^*)&=&3\,{\cal R}&\quad&H^1(L_4\otimes L_1^*)&=&3\,{\cal R}&\quad& H^1(L_2\otimes L_3^*)&=&5\,{\cal R}\\
 H^1(L_4\otimes L_2^*)&=&{\cal R}&\quad&H^1(L_2\otimes L_5^*)&=&3\,{\cal R}\;,&&
 \end{array}
\end{equation} 
where
\be
 \tilde{\cal R}=  (0,1)\oplus (1,0)\oplus (1,1)
\ee 
is the regular representation minus the trivial one.

The Wilson line is specified by a group homomorphism from $\Gamma=\mathbb{Z}_2\times\mathbb{Z}_2$ into hypercharge and it can be represented by two irreducible $\mathbb{Z}_2\times\mathbb{Z}_2$ representations $W_2$ and $W_3$. For the present model we choose $W_2=(0,1)$ and $W_3=(0,0)$, so that the  $\mathbb{Z}_2\times\mathbb{Z}_2$ charges of the particles are
\begin{equation}\label{Wf}
\begin{array}{lllll}
  W(d)=W_3^*=(0,0)&\quad&W(L)=W_2^*=(0,1)&\quad&W(Q)=W_2\otimes W_3=(0,1)\\
  W(u)=W_3\otimes W_3=(0,0)&\quad&W(e)=W_2\otimes W_2=(0,0)&\quad&W(H)=W_2^*=(0,1)\\
  W(T)=W_3^*=(0,0)&\quad&W(\bar{H})=W_2=(0,1)&\quad&W(\bar{T})=W_3=(0,0)\; .
\end{array}  
\end{equation}
 Here, we have used the obvious notation for the physical particles and $T$, $\bar{T}$ denote the Higgs triplets. To find the number of physical particles $f$ with $\mathbb{Z}_2\times\mathbb{Z}_2$ charge $W(f)$ from a given cohomology $H^1(X,{\cal L})$ we have to extract the $\mathbb{Z}_2\times\mathbb{Z}_2$ singlets from $H^1(X,{\cal L})\otimes W(f)$. From Eqs.~\eqref{lbcohgrad}, \eqref{Wf} and the identification of cohomologies and particles discussed below Eq.~\eqref{lbcoh} this leads to the spectrum
 \begin{equation}
  2\,{\bf 10}_2\,,\;{\bf 10}_4\,,\;2\,\bar{\bf 5}_{4,5}\,,\;\bar{\bf 5}_{2,5}\,,\;H_{2,4}\,,\;\bar{H}_{2,4}\,,\;3\,{\bf 1}_{2,1}\,,\;3\,{\bf 1}_{4,1}\,,\;
  5\,{\bf 1}_{2,3}\,,\;3\,{\bf 1}_{2,5}\,,\;{\bf 1}_{4,3}\; . \label{SMAb}
\end{equation}  
For ease of notation we have written the families in GUT notation but we should think of these as being broken up into standard model multiplets. The above spectrum is a precise MSSM spectrum plus a number of bundle moduli singlets. It should be compared with the GUT spectrum~\eqref{AbSU(5)spec} at the Abelian locus. All chiral parts of this spectrum have been divided by four, the order of our $\mathbb{Z}_2\times\mathbb{Z}_2$  symmetry, as expected. From the three vector-like pairs, $\bar{\bf 5}_{2,4}$--${\bf 5}_{2,4}$, we have removed all triplets and kept only the two Higgs doublets $H_{2,4}$ and $\bar{H}_{2,4}$. This  works because the relevant cohomologies $H^1(X,L_2\otimes L_4)=\tilde{\cal R}$ and $H^1(X,L_2^*\otimes L_4^*)= \tilde{\cal R}$ are missing the trivial $\mathbb{Z}_2\times\mathbb{Z}_2$ representation. Hence, choosing $W_3=(0,0)$ projects out all the triplets and $W_2=(0,1)$ selects precisely one of the doublets from each cohomology. 
\subsection{The equivariant structure for the non-Abelian bundle}
To work out the standard model away from the Abelian locus we need to find a suitable equivariant structure on the non-Abelian bundle $V=U\oplus L$. For the line bundle $L$ we choose the equivariant structure corresponding to the trivial $\mathbb{Z}_2\times\mathbb{Z}_2$ representation, so
\begin{equation}
 L={\cal O}_X(1,1,-1,-1)_{(0,0)}\; .
\end{equation} 
Note that this is the same choice as for $L=L_5$ at the Abelian locus, see Eq.~\eqref{lbequiv}. It remains to find an equivariant structure on the monad bundle $U$. The first step is to assign equivariant structures to the line bundle sums $B$ and $C$ used to define the monad bundle~\eqref{monaddef}. To this end, we choose the following equivariant structures on $B$ and $C$:
\bea 
B& = &{\cal O}_X (-1, 0, 0, 1)_{(0, 0)} \oplus  {\cal O}_X (-1, -1, 2, 2)_{(1, 0)} \oplus {\cal O}_X (-1, -1, 2, 2)_{(0, 0)} 
\nonumber \\
&& \oplus\; {\cal O}_X (0, 1, -1, 0)_{(0, 0)} \oplus  {\cal O}_X (1, 1, 0, 0)_{(1, 0)} \oplus {\cal O}_X (1, 1, 0, 0)_{(0, 0)}\label{Bequiv}\\
C&=& {\cal O}_X (-1, 1, 2, 2)_{(1, 0)} \oplus  {\cal O}_X (1, 1, 0, 2)_{(1, 0)}\,. \label{Cequiv}
\eea
The next step is to restrict the map $f$ in~\eqref{eq:monadmap} to be consistent with the above charge assignments. Denoting the homogeneous coordinates 
on ${\mathbb P}^1 \times {\mathbb P}^1 \times {\mathbb P}^1 \times{\mathbb P}^1$ by $(t_0, t_1)$, $(x_0, x_1)$, $(y_0, y_1)$, $(z_0, z_1)$, respectively, the generic form of this restricted map becomes
\begin{equation}\label{fequiv}
 \begin{array}{lll}
f_{(0, 1, 2, 1)}&=& a_1 y_0 y_1 (x_0 z_0 + x_1 z_1) + a_2 (y_0^2 + y_1^2) (x_0 z_1 + x_1 z_0) + a_3 (y_0^2 - y_1^2) (x_0 z_1 - x_1 z_0)\\
f_{(2, 1, 0, 1)}&=& b_1 t_0 t_1 (x_0 z_0 + x_1 z_1) + b_2 (t_0^2 + t_1^2) (x_0 z_1 + x_1 z_0) + b_3 (t_0^2 - t_1^2) (x_0 z_1 - x_1 z_0)\\
 f_{(1, 0, 1, 2)}&=& c_1 z_0 z_1 (t_0 y_0 + t_1 y_1) + c_2 (z_0^2 + z_1^2) (t_0 y_1 + t_1 y_0) + c_3 (z_0^2 - z_1^2) (t_0 y_1 - t_1 y_0)\\
 f_{(0, 2, 0, 0)} &=& d_1 (x_0^2 + x_1^2)\,, \quad f_{(0, 2, 0, 0)}^{\prime} = d_2 x_0  x_1\\
 f_{(0, 0, 0, 2)} &=& d_3 (z_0^2 + z_1^2)\,, \quad f_{(0, 0, 0, 2)}^{\prime} = d_4 z_0  z_1\; .
 \end{array}
 \end{equation}
 where $a_1, b_1,\ldots$  are arbitrary coefficients. We have checked that with $f$ restricted in this way the monad $U$ is indeed a rank four vector bundle (rather than a sheaf) and, following the same steps as in Ref.~\cite{Buchbinder:2013dna}, we have also verified that $V=U\oplus L$ remains supersymmetric in the cone~\eqref{CU}. To arrive at the above choice of equivariant structure we have been guided by two requirements. First, the restricted monad map should still be sufficiently general for $U$ to remain a bundle. In fact, this requirement excludes choosing the trivial equivariant structure for all line bundles in $B$ and $C$. Secondly, at the split locus, the equivariant structure for $V$ should coincide with the one we have made for the line bundle sum $V_0$ in Eq.~\eqref{lbequiv}.
 
As at the Abelian locus, we choose the Wilson line $W_2=(0,1)$ and $W_3=(0,0)$. This leads to the Wilson line charges of the various multiplets as in Eq.~\eqref{Wf}. To find the downstairs spectrum we need to work out the $\mathbb{Z}_2\times\mathbb{Z}_2$ representation content of the various cohomologies. Let us discuss in some detail the fate of the cohomology $H^1(X,U)$ which gives rise to the $12\,{\bf 10}_{-1}$ families upstairs. The long exact sequence associated to the monad sequence~\eqref{monaddef} reads
 \be 
 0 \longrightarrow H^0 (X, B) \longrightarrow H^0 (X, C) \longrightarrow H^1 (X, U) \longrightarrow H^1 (X, B) \longrightarrow 0\,, 
 \ee
so that  
\be
 H^1(X,U)\cong {\rm Coker}(H^0(X,B)\rightarrow H^0(X,C))\oplus H^1(X,B)\; .
\ee 
For the $\mathbb{Z}_2\times\mathbb{Z}_2$ representation content of the various line bundle sums we find
\begin{equation}
 H^0(X,B)=2\,{\cal R}\;,\quad H^0(X,C)=3\,{\cal R}\;,\quad H^1(X,B)=2\,{\cal R}\quad\Longrightarrow\quad H^1(X,U)=3\,{\cal R}\; .
\end{equation} 
For a given charge of a standard model multiplet contained in ${\bf 10}$, as in Eq.~\eqref{Wf}, precisely one representation from the regular one, ${\cal R}$, is selected and we remain with three families of matter. 
The remaining parts of the chiral spectrum can be worked out in a similar way. Specifically, we find
\begin{equation}
 H^1((X, U\otimes L)= 3\,{\cal R}\;,\quad H^1(X,U\otimes L^*)=3{\cal R}\; .\label{Hequiv}
\end{equation} 
After multiplying with the relevant Wilson line charges~\eqref{Wf} and projecting onto the singlets this leads to three $\bar{\bf 5}_3$ families and three singlet (right-handed neutrinos) ${\bf 1}_{-5}$.

Obtaining the spectrum in the Higgs sector is substantially more complicated. The crucial result, shown in Appendix B, is that
\be 
H^1(X, \wedge^2 U)= \tilde{\cal R}\; , \quad H^1(X,\wedge^2U^*)=\tilde{\cal R}\; ,
\ee
where $\tilde{\cal R}=(0,1)\oplus (1,0)\oplus (1,1)$. Hence, our choice of $W_3=(0,0)$ removes all Higgs triplets and $W_2=(0,1)$ implies that only one Higgs doublet each is kept. Note that this result is consistent with the representation structure found at the Abelian locus where the relevant cohomologies, $H^1(X,L_2\otimes L_4)$ and $H^1(X,L_2^*\otimes L_4^*)$ in Eq.~\eqref{lbcohgrad} are given by the same representation $\tilde{\cal R}$. This is not surprising since we have chosen the equivariant structures to be compatible. In summary, the spectrum charged under the gauge group $SU(3)\times SU(2)\times U_Y(1)\times U_X(1)$ is given by
\begin{equation}
 3\,{\bf 10}_{-1}\,,\;3\,\bar{\bf 5}_3\,,\;H_{-2}\,,\;\bar{H}_2\,,\; 3\,{\bf 1}_{-5}\; ,\label{SMspec}
\end{equation} 
that is, an MSSM spectrum plus three right-handed neutrinos. As before, the GUT multiplets should be thought of as being broken up into their standard model components. In addition, we expect nine uncharged singlets, ${\bf 1}_0$. This can be inferred from the spectrum~\eqref{SMAb} at the Abelian locus where we have a total of $15$ singlets. As we move away from the Abelian locus, three of those will become massive as the three $U(1)$ symmetries are broken while the $3\,{\bf 1}_{2,5}$ are identified with the $3\, {\bf 1}_{-5}$ in \eqref{SMspec}. This leaves nine uncharged singlets ${\bf 1}_0$. Combining the standard hypercharge with $U_X(1)$ as in Eq.~\eqref{Qdef} leads to the standard $B-L$ charge for all fields. This is of course expected as we have constructed a regular model.

The calculation of the equivariant cohomology of $H^1 (X, U \otimes U^*)$ is quite involved and will not be presented in this paper. 


\subsection{Comments on extensions and topological transitions}


Having constructed a vector bundle $V=U \oplus L$ with structure group $S(U(4) \times U(1))$ which is polystable at the stability wall 
\be
\mu (L)= \vec{s} \cdot (1, 1, -1, -1)=0
\label{4.1}
\ee
we can ask a if it can be extended to a full $SU(5)$ bundle\footnote{For simplicity, we present our discussion at the upstairs level. All our conclusions remain the same for the bundle on the quotient space.}. Considering the GUT spectrum~\eqref{AbSU(5)spec} at the Abelian locus, the only singlets which connect $L=L_5$ with any of the other line bundles are ${\bf 1}_{2,5}$. Hence, our $S(U(4)\times U(1))$ bundles can be thought off as a deformation of the model at the Abelian locus with VEVs for all singlets except for the $12\, {\bf 1}_{2,5}$ switched on. At a generic $S(U(4)\times U(1))$ locus, the $12 \, {\bf 1}_{2,5}$ singlets become $12\, {\bf 1}_{-5}$, as comparison with spectrum~\eqref{NonAbSU(5)spec} shows, and switching on VEVs for these fields as well deforms the structure group further to $SU(5)$.

Mathematically, this can be described by the extension sequence
\be 
0 \longrightarrow U \longrightarrow V' \longrightarrow L \longrightarrow 0\,. 
\label{4.2}
\ee
for the bundle $V'$. The moduli space of these bundles is given by ${\rm Ext}^1(L, U) = H^1 (X, U \otimes L^*)$ which is precisely the cohomology containing the $12$ singlets ${\bf 1}_{-5}$. At the origin in ${\rm Ext}^1(L, U)$, that is, for vanishing ${\bf 1}_{-5}$ singlet VEVs, the extension is trivial so $V'=U\oplus L$, while for non-vanishing elements of ${\rm Ext}^1(L, U)$, corresponding to non-vanishing ${\bf 1}_{-5}$ singlet VEVs, $V'$ becomes a non-trivial extension with an $SU(5)$ structure group.

We can also ask about the opposite extension
\be
0 \longrightarrow L \longrightarrow V^{\prime \prime} \longrightarrow U \longrightarrow 0\,.
\label{4.3}
\ee
whose moduli space is governed by ${\rm Ext}^1(U, L) =H^1 (X, U^* \otimes L)$. A non-trivial extension of this kind corresponds to an $SU(5)$ bundle $V^{\prime\prime}$ topologically different from $V^\prime$ and the stability wall~\eqref{4.1} marks the topological transition between these two $SU(5)$ bundles. However, in the present case, $h^1(X,U^*\otimes L)=0$, that is there are no right-handed anti-neutrinos ${\bf 1}_5$ in the spectrum~\eqref{NonAbSU(5)spec}. This means that non-trivial extension bundles $V^{\prime\prime}$ do not exist for our example.  Hence, the locus~\eqref{4.1} is a genuine stability wall for the bundle $V'$ beyond which it cannot be extended in a supersymmetric way.

A crucial feature of our $S(U(4)\times U(1))$ model is that the vector-like ${\bf 5}$--$\bar{\bf 5}$ pairs which lead to the Higgs multiplet remain massless. This can also be understood from the GUT spectrum~\eqref{NonAbSU(5)spec}. The only allowed coupling which might give rise to a mass term is of the form ${\bf 1}_{-5}\,{\bf 5}_2\,\bar{\bf 5}_3$. However, as long as we keep the VEV of ${\bf 1}_{-5}$ zero, as we do at the $S(U(4)\times U(1))$ locus, no mass is generated. This also suggests that away from the $S(U(4)\times U(1))$ locus, when we switch on ${\bf 1}_{-5}$ VEVs and the structure group becomes $SU(5)$, the vector-like ${\bf 5}$--$\bar{\bf 5}$ pairs do become massive and are removed from the low-energy spectrum. This can indeed be confirmed by a calculation of the bundle cohomology for $V'$. Hence, we can only expect massless Higgs doublets at the $S(U(4)\times U(1))$ locus and the model is not phenomenologically viable away from it.


\section{Proton stability}\label{sec:proton}
\label{sec:proton}
Probably the most important phenomenological constraint on string models, beyond obtaining the correct spectrum of low-energy particles, arises from proton stability. We would now like to discuss proton stability for our standard model. For simplicity, we will carry this discussion out in the GUT version of the model but the discussion is completely analogous - and leads to the same conclusion - for the associated standard model. 

Dimension four operators which can lead to proton decay are of the form ${\bf 10}\,\bar{\bf 5}\,\bar{\bf 5}$ in GUT language. A quick glance at the GUT spectrum~\eqref{NonAbSU(5)spec} shows that such operators are forbidden by the $U_X(1)$ symmetry or, equivalently, the $B-L$ symmetry present in the associated standard model. It is of course well-known that a $B-L$ symmetry forbids these dimension four operators in the superpotential so this does not come as a surprise. At any rate, we conclude that our model is safe from proton decay induced at the level of dimension four operators. 

What about dimension five operators? Considering the spectrum~\eqref{NonAbSU(5)spec}, proton-decay inducing dimension five operators of the from ${\bf 10}_{-1}\,{\bf 10}_{-1}\,{\bf 10}_{-1}\,\bar{\bf 5}_3$ are allowed by the $U_X(1)$ symmetry. Again, this is expected since operators of this type are well-known to be consistent with $B-L$. So it appears that our model has a problem with proton decay induced by dimension five operators.

This is where the existence of the Abelian locus in the moduli space of our model becomes important. At the Abelian locus the gauge symmetry is enhanced from $SU(5)\times U_X(1)$ to $SU(5)\times S(U(1)^5)$. A glance at the spectrum~\eqref{AbSU(5)spec} shows that operators ${\bf 10}\,{\bf 10}\,{\bf 10}\,\bar{\bf 5}$ are not invariant under the $S(U(1)^5)$ symmetry. What is more, all such operators with any number of additional singlet insertions, so operators of the form ${\bf 1}_{{\bf q}_1}\cdots {\bf 1}_{{\bf q}_n}\,{\bf 10}\,{\bf 10}\,{\bf 10}\,\bar{\bf 5}$ are also forbidden, given the available charges ${\bf q}_i$ of singlet fields in \eqref{AbSU(5)spec}. 

As we have discussed earlier, switching on singlet field VEVs corresponds to moving away from the Abelian locus to a model with non-Abelian structure group and gauge group $SU(5)\times U_X(1)$. Hence, the absence of all proton-decay inducing dimension five operators at the Abelian locus, including those with an arbitrary number of singlet insertions, means that these operators remain forbidden even away from the Abelian locus.

We conclude that our model, both at the Abelian locus and away from it, is safe from fast proton decay induced by dimension four and five operators. From the viewpoint of the $SU(5)\times U_X(1)$ model the absence of dimension five operators is unexpected since it is not enforced by any apparent low-energy symmetries of this model. Also, we are not aware of a method by which the coefficient of these dimension five operators can be calculated directly. Instead, their absence is inferred indirectly from the existence of the Abelian locus where the  symmetry is enhanced. 

\section{Conclusions}
In this paper, we have considered heterotic Calabi-Yau models with bundle structure group $S(U(4)\times U(1))$. We have seen that for a certain subclass of  ``regular" such models, the additional $U(1)$ gauge symmetry present at low energy can combine with hypercharge to a standard $B-L$ symmetry. However, unlike in heterotic models with a rank four vector bundle and an underlying $SO(10)$ GUT theory, the $U(1)$ vector boson is super-massive and not of phenomenological concern. 

We have studied a particular example of a regular model, based on the tetra-quadric Calabi-Yau manifold. This model has a perfect MSSM spectrum plus some additional (bundle moduli) fields which are uncharged under the standard model group. Due to the $B-L$ symmetry, dimension four operators leading to fast proton decay are forbidden. In addition, dimension five proton-decay inducing operators are also absent, essentially due to the presence of a locus in moduli space with enhanced symmetry. At this locus, the bundle structure group becomes Abelian and the low-energy symmetry enhances by three $U(1)$ factors. These additional $U(1)$ symmetries forbid all relevant dimension five operators, including those with bundle moduli singlet insertions. This means that, even when moving away from the special locus by switching on bundle moduli VEVs, the dangerous dimension five operators cannot be generated perturbatively.\footnote{Our discussion did not take into account non-perturbative effects. In general, couplings that are forbidden by the $U(1)$~symmetries at the perturbative level, can in principle be generated non-perturbatively. However, these would be highly suppressed and hence would not affect the present work.}

More generally, we have developed some of the methods required to deform heterotic line bundle models to models with non-Abelian bundle structure group, thereby exploring the full bundle moduli space. Heterotic line bundle models are relatively easy to construct but usually reside in a larger moduli space of non-Abelian bundles. A serious phenomenological analysis of heterotic line bundle models requires an understanding of this larger moduli space and the present paper is laying some of the required groundwork. Some of the present results should generalized to the whole class of line bundle standard models~\cite{Anderson:2011ns, Anderson:2012yf,Linebundles} and we hope to return to this problem in a future publication.


\section*{Acknowledgements} 

We would like to thank Xenia de la Ossa, Graham Ross and Daniel Waldram for insightful discussions on the material presented here.
The work of  E.~I.~B. is supported by the ARC Future Fellowship FT120100466.
E.~I.~B. would like to thank physics department at Oxford University where some of this work was done for warm hospitality. A.~C.'s work is supported by STFC.  A.~L.~is partially supported by the EPSRC network grant EP/l02784X/1 and by the STFC consolidated grant ST/L000474/1.


\newpage
\appendix
\renewcommand\baselinestretch{1.1}
\section{Identities for characteristic classes and indices}\label{app:identities}
In this Appendix we collect a number of standard identities for characteristic classes which are useful for the split bundles $U\oplus L$ used in the main part of the paper.

Let  $U$, $W$ be vector bundles and $L$ be a line bundle on a Calabi--Yau threefold $X$. The Chern characters of $U$, $W$ satisfy the following properties 
\be 
\ch (U \oplus W) = \ch (U) + \ch (W)\,, \quad \ch (U \otimes W) = \ch (U) \wedge \ch (W)
\label{A1}
\ee
and they are related to the Chern classes by 
\begin{equation}\label{A2}
\begin{array}{lllllll}
\ch_0 (U)& =&{\rm rk} (U)&\quad&\ch_1 (U)&=& {\rm c}_1 (U)\\
\ch_2 (U) &=&-{\rm c}_2 (U) +\frac{1}{2}{\rm c}_1 (U)^2&\quad&
\ch_3 (U)& =&\frac{1}{2} {\rm c}_3 (U) -\frac{1}{2}{\rm c}_1 (U)\, {\rm c}_2 (U) +\frac{1}{6}{\rm  c}_1 (U)^3\; .
\end{array}
\end{equation}
For a line bundle $L$ the total Chern class is $ c (L)= 1+ c_1 (L)$ and, hence, 
\begin{equation}
\ch_0 (L^p)= 1\,, \quad \ch_2 (L^p)= p\, {\rm c}_1 (L) \,, \quad
\ch_2 (L^p)= \frac{p^2}{2} {\rm c}_1 (L)^2\,, \quad \ch_3 (L^p)= \frac{p^3}{6}  {\rm c}_1 (L)^3 \,, 
\label{A3}
\end{equation}
where $p$ is an arbitrary integer. The Chern character of the tensor product $U \otimes L$ can be decomposed as 
\bea
&& 
\ch_0 (U \otimes L) = {\rm rk} (U)\,, 
\nonumber\\
&& 
\ch_1 (U \otimes L) = \ch_1 (U) + \rk (U)\, \ch_1 (L) \,, 
\nonumber \\
&& 
\ch_2 (U \otimes L) = \ch_2 (U) + \rk (U)\, \ch_2 (L) +\ch_1 (U)\, \ch_1 (L)\,, 
\nonumber \\
&& 
\ch_3 (U \otimes L) = \ch_3 (U) + \rk (U)\, \ch_3 (L) +\ch_2 (U)\, \ch_1 (L) + \ch_1 (U)\, \ch_2 (L)\,. 
\label{A4}
\eea
If $U$ is a rank 4 vector bundle the characteristic classes of the second wedge power $\wedge^2 U$ are given by 
\begin{equation}
\begin{array}{lllllll}
{\rm c}_0 (\wedge^2 U)&=& 1&\quad&\ch_0 (\wedge^2 U)&=& 6\\
{\rm c}_1 (\wedge^2 U)&=& 3\, {\rm c}_1 (U)&\quad& \ch_1 (\wedge^2 U)&=&  3\, \ch_1 (U)\\
{\rm c}_2 (\wedge^2 U)&=&  2\, {\rm c}_2 (U) + 3\, {\rm c}_1 (U)^2&\quad&\ch_2 (\wedge^2 U)&=& 2\, \ch_2 (U) + \frac{1}{2}  \ch_1 (U)^2\\
{\rm c}_3 (\wedge^2 U)&=&4\, {\rm c}_2 (U)\, {\rm c}_1 (U)  +  {\rm c}_1 (U)^3&\quad& \ch_3 (\wedge^2 U)&=& \ch_1 (U)\, \ch_2 (U)
\end{array}
\label{A5}
\end{equation}
The index of a vector bundle $U$ is given by 
\be 
{\rm Ind} (U)= \int_X \ch (U)\,{\rm Td} (TX)\,, 
\label{A6}
\ee
with the Todd class of a Calabi-Yau three-fold given by
\be 
{\rm Td} (TX) = 1+\frac{1}{12} {\rm c}_2 (TX) \; .
\label{A6.1}
\ee
Using eqs.~\eqref{A2}, \eqref{A3}, \eqref{A5} -- \eqref{A6.1} we obtain the following expressions for the indices of $L$ and $\wedge^2 U$
\begin{equation}
{\rm Ind} (L )= \frac{1}{12} \left( 2 {\rm c}_1 (L)^3 + {\rm c}_1 (L)\, {\rm c}_2 (TX)\right) \,, \quad
{\rm Ind} (\wedge^2 U )=  \ch_1 (U)\, \ch_2 (U) +\frac{1}{4} \ch_1 (U)\, {\rm c}_2 (TX)\,. 
\end{equation}
while the index of $U\otimes L$ is given by
\begin{equation}
 \ind(U\otimes L)=\ind(U)+4\ind(L)-\frac{1}{2}{\rm c}_1(L)^3+{\rm c}_1(L)\ch_2(U)\; .
\end{equation}

\renewcommand\baselinestretch{.9}
\section{The equivariant structure and the Higgs sector}
In this Appendix we describe the calculation leading to the result~\eqref{Hequiv} for the equivariant cohomology in the Higgs sector which is crucial to show that we retain a pair of Higgs doublets and remove all Higgs triplets. First we recall that the monad bundle $U$ is defined by the short exact sequence~\eqref{monaddef} and the $\mathbb{Z}_2\times\mathbb{Z}_2$ equivariant structure on $U$ is defined by the equivariant structures~\eqref{Bequiv}, \eqref{Cequiv} on the underlying line bundle sums $B$, $C$, together with the restricted monad map~\eqref{fequiv}.

The down Higgs doublet arises from the cohomology $H^1(X,\wedge^2 U)$ while the up Higgs is contained in $H^1(X,\wedge^2 U^*)$. We will focus on the former and determine the $\mathbb{Z}_2\times\mathbb{Z}_2$ representation for $H^1(X,\wedge^2 U)$. The representation of $H^1(X,\wedge^2 U^*)$ is simply the dual of this representation.

We begin with the second wedge power sequence
\begin{equation}
 0\rightarrow \wedge^2U\rightarrow \wedge^2 B\rightarrow B\otimes C\rightarrow S^2C\rightarrow 0
\end{equation}
associated to the monad sequence~\eqref{monaddef}. Splitting this up into two short exact sequences by introducing the co-kernel $K$ gives
\begin{equation}
 \begin{array}{cccccccccccc} 
  &\wedge^2U&\rightarrow&\wedge^2B&\rightarrow&K&\quad\quad&K&\rightarrow&B\otimes C&\rightarrow&S^2C\\
h^0(X,\cdot)&0&&53&&56&&56&&150&&96\\
&&&(14,12,15,12)&&&&&&(38,37,38,37)&&(24,24,24,24)\\
h^1(X,\cdot)&3&&85&&88&&88&&134&&48\\
&&&(22,20,23,20)&&&&&&(34,33,34,33)&&(12,12,12,12)\\
h^2(X,\cdot)&3&&0&&0&&0&&0&&0\\
h^3(X,\cdot)&0&&0&&0&&0&&0&&0\\
 \end{array} 
\end{equation}
Here, the integers indicate the cohomology dimensions in the associated long exact sequences and, for the line bundle sums, the four-vectors underneath provide the breakdown of these cohomologies into the four irreducible $\mathbb{Z}_2\times\mathbb{Z}_2$ representations, in the order $(0,0),\,(0,1),\,(1,0),\,(1,1)$. We know that
\vspace{-4pt}
\begin{eqnarray}
 H^1(X,\wedge^2U)&\cong& {\rm Coker}(H^0(X,\wedge^2B)\rightarrow H^0(X,K))\\
 H^0(X,K)&\cong&{\rm Ker}(H^0(X,B\otimes C)\rightarrow H^2(X,S^2C))\; .
\vspace{-4pt}
\end{eqnarray}
Further, we introduce the co-kernel\vspace{-8pt}
\begin{equation}
 {\cal C}={\rm Coker}(H^0(X,B\otimes C)\rightarrow H^0(X,S^2C))
\vspace{-4pt}
\end{equation}
whose dimension is ${\rm dim}({\cal C})=2$. From now on, we will consider cohomology dimensions as being broken up into multiplicities of the four  $\mathbb{Z}_2\times\mathbb{Z}_2$ representations so we write ${\bf c}=(c_1,c_2,c_3,c_4)={\rm dim}({\cal C})$, where $\sum_{i=1}^4c_i=2$. 
Then, combining the above equations we find for the dimension of the Higgs cohomology that
\vspace{-4pt}
\begin{equation}
 h^1(\wedge^2 U)=h^0(B\otimes C)-h^0(S^2C)-h^2(\wedge^2 B)+{\bf c}=(c_1,1+c_2,c_3-1,c_4)\; . \label{h1res}
\vspace{-4pt}
\end{equation}
To complete the calculation we need to work out the $\mathbb{Z}_2\times\mathbb{Z}_2$ representation content of the co-kernel ${\cal C}$. This involves explicitly constructing the $150$-dimensional space $H^0(X,B\otimes C)$, the $96$-dimensional space $H^0(X,S^2C)$ and the relevant map, induced by the specialized monad map~\eqref{fequiv}, between those spaces. Using computer algebra this leads to ${\bf c}=(0,0,2,0)$. We note that this result is consistent with the constraint $c_3\geq 1$ implied by Eq.~\eqref{h1res}. Inserting into Eq.~\eqref{h1res} then gives
\begin{equation}
 h^1(X,\wedge^2U)=(0,1,1,1)\quad\Longrightarrow\quad H^1(X,\wedge^2 U)=\tilde{\cal R}\; ,
\end{equation}
where $\tilde{\cal R}=(0,1)\oplus (1,0)\oplus (1,1)$.  


\newpage
\bibliography{bibfile}{}
\bibliographystyle{utcaps}

\end{document}